\pdfoutput=1
\documentclass[twocolumn,english,aps,prb,showpacs,superscriptaddress]{revtex4}
\usepackage{ae,aecompl}
\usepackage[T1]{fontenc}
\usepackage{color}
\usepackage{amstext}
\usepackage{amssymb}
\usepackage{graphicx}

\makeatletter

\newcommand{\lyxmathsym}[1]{\ifmmode\begingroup\def\b@ld{bold}
  \text{\ifx\math@version\b@ld\bfseries\fi#1}\endgroup\else#1\fi}

\@ifundefined{textcolor}{}
{%
 \definecolor{BLACK}{gray}{0}
 \definecolor{WHITE}{gray}{1}
 \definecolor{RED}{rgb}{1,0,0}
 \definecolor{GREEN}{rgb}{0,1,0}
 \definecolor{BLUE}{rgb}{0,0,1}
 \definecolor{CYAN}{cmyk}{1,0,0,0}
 \definecolor{MAGENTA}{cmyk}{0,1,0,0}
 \definecolor{YELLOW}{cmyk}{0,0,1,0}
 }

\makeatother

\usepackage{babel}
\begin{document}

\title{Stability of the Ni sites across the pressure-induced metallization
in $YNiO_{3}$}

\author{Aline Y. Ramos}

\email{aline.ramos@grenoble.cnrs.fr}

\selectlanguage{english}%

\affiliation{Institut Néel, CNRS et Université Joseph Fourier, BP 166, F-38042
Grenoble Cedex 9, France}

\author{Cinthia Piamonteze}

\affiliation{Swiss Light Source, Paul Scherrer Institut, CH-5232 Villigen PSI,
Switzerland}

\author{Hélio C. N. Tolentino}

\affiliation{Institut Néel, CNRS et Université Joseph Fourier, BP 166, F-38042
Grenoble Cedex 9, France}

\author{Narcizo M. Souza-Neto}

\affiliation{Laboratório Nacional de Luz Síncrotron - P.O. Box 6192, 13084-971,
Campinas, São Paulo, Brazil}

\author{Oana Bunau}

\affiliation{Institut Néel, CNRS et Université Joseph Fourier, BP 166, F-38042
Grenoble Cedex 9, France}

\author{Yves Joly}

\affiliation{Institut Néel, CNRS et Université Joseph Fourier, BP 166, F-38042
Grenoble Cedex 9, France}

\author{Stéphane Grenier}

\affiliation{Institut Néel, CNRS et Université Joseph Fourier, BP 166, F-38042
Grenoble Cedex 9, France}

\author{Jean-Paul Itié}

\affiliation{Synchrotron SOLEIL, L'Orme des Merisiers, Saint-Aubin, BP 48, 91192
Gif-sur-Yvette Cedex, France}

\author{Néstor E. Massa}

\affiliation{Laboratorio Nacional de Investigación y Servicios en Espectroscopía
óptica-Centro CEQUINOR, Universidad Nacional de La Plata, C.C. 962,
1900 La Plata, Argentina}

\author{José A. Alonso}

\affiliation{Instituto de Ciencia de Materiales de Madrid, Cantoblanco, E-28049
Madrid, Spain}

\author{Maria J. Martinez-Lope}

\affiliation{Instituto de Ciencia de Materiales de Madrid, Cantoblanco, E-28049
Madrid, Spain}
\begin{abstract}
The local environment of nickel atoms in $YNiO_{3}$ across the pressure-
induced insulator to metal (IM) transition was studied using X-ray
absorption spectroscopy (XAS) supported by \emph{ab initio} calculations.
The monotonic contraction of the $NiO_{6}$ units under applied pressure
observed up to 13 GPa, stops in a limited pressure domain around 14
GPa, before resuming above 16~GPa. In this narrow pressure range,
crystallographic modifications basically occur in the medium/long
range, not in the $NiO_{6}$ octahedron, whereas the evolution of
the near-edge XAS features can be associated to metallization. \emph{Ab
initio} calculations show that these features are related to medium
range order, provided that the Ni-O-Ni angle enables a proper overlap
of the Ni $e_{g}$ and O $2p$ orbitals. Metallization is then not
directly related to modifications in the average local geometry of
the $NiO_{6}$ units but more likely to an inter-octahedra rearrangement.
These outcomes provides evidences of the bandwidth driven nature of
the IM transition.
\end{abstract}


\maketitle

\section{Introduction}

The \emph{Ni} perovskite family $RNiO_{3}$ (\emph{$R=Y$} and rare-earth
\emph{$\neq La$}) presents a localized 3d-electron behavior at low
temperatures and a first-order insulator to metal (IM) transition
as temperature increases\cite{Garcia-Munoz-PRB92,Torrance-PRB92,Catalan-PhaseTrans08}.
The transition temperature ($T_{IM}$) is determined by the \emph{R
}ionic radius that modifies the mismatch between the Ni-O and $R$-O
bond lengths. Such modification goes along with changes in the Ni-O-Ni
superexchange angle. A decrease in $T_{IM}$ reflects an increase
in the Ni $3d\, e_{g}$ and O $2p$ orbitals hybridization, concomitant
with an increase in the band-width \cite{Zhou-PRB00}. In compounds
with small \emph{R} ions (Y, Ho, Er, Tm, Yb, and Lu) the crystallographic
symmetry changes from monoclinic to orthorhombic across the IM transition\cite{Alonso-PRL99,Alonso-PRB01,Medarde-PRB09}.
The single $Ni^{3+}$ site in the metallic phase split in the insulating
phase into two nonequivalent $Ni1$ and $Ni2$ sites, with slightly
different average Ni-O distances. These two different average distances
are interpreted as a signature of charge order (or charge disproportionation)\cite{Alonso-PRL99,Mazin-PRL07,Staub-PRL02}.
It has also been proposed that the difference in \emph{Ni} local environment
reflects the presence of an ionic bonding at Jahn\textendash{}Teller
(\emph{JT}) distorted $Ni2$ (larger) sites and covalent bonding at
$Ni1$ (smaller) sites, with possible dynamical fluctuations between
both sites \cite{Zhou-PRB04b,Cheng-PRB10}. Confirming such fluctuations,
a strong electron-lattice coupling in a \textit{JT}-distorted lattice
has been evidenced in small-R compounds\cite{Medarde-PRL98}. For
the largest-$R$ compounds the time scale of the fluctuations is expected
to be shorter and the splitting is hardly observable by elastic scattering
techniques\cite{Staub-PRL02,Zhou-PRB04b}, leading to the description
of an average structure. Changes in the local symmetry can be tracked
by X-ray absorption spectroscopy (XAS). 

As a structural technique able to probe dynamical changes, XAS has
shown\textit{\emph{ that the two Ni sites coexist in both insulating
and metallic state in several $RNiO_{3}$ compounds \cite{Piamonteze-PRB05a,Piamonteze-NIMB06}.}}
In $PrNiO_{3}$ local symmetry changes accompanying electronic delocalization
were found incompatible with the average long range order proposed
from X-ray and neutron diffraction \cite{Acosta-PRB08}. These studies
emphasize the existence of an inhomogeneous structure at the local
scale and suggest a common behavior in all $RNiO_{3}$ compounds for
the local electronic and magnetic state \cite{Piamonteze-PRB05a}. 

Among the~$RNiO_{3}$ compounds, $YNiO_{3}$ presents one of the
largest monoclinic distortion at room temperature, due to the small
size of the $Y^{3+}$~ions\cite{Alonso-PRB01}. The thermal-induced
metallic phase occurs simultaneously with the vanishing of the long-range
monoclinic distortion~\cite{Alonso-PRL99}. However, \textit{\emph{the
two Ni sites coexistence, clearly observed by XAS even in the orthorhombic
phase \cite{Piamonteze-PRB05a,Piamonteze-NIMB06}, supports a model
of stable short-range scale distortion. }}

The application of an external pressure provides a unique tool to
further investigate the relationship between structural distortions
and electronic properties. Hydrostatic pressure reduces the unit cell
volume and shrinks Ni-O bond lengths while straightening the bond
angle and stabilizing the metallic phase \cite{Canfield-PRB98,Garcia-Munoz-PRB04,Zhou-PRB00,Obradors-PRB93}.
In $YNiO_{3}$ Garcia Munoz and coworkers reported by X-ray diffraction
a sudden structural modification around 14 GPa, consistent with a
monoclinic to orthorhombic transition\cite{Garcia-Munoz-PRB04}. At
this pressure, an increase in electronic conductivity and a phonon
screening are also observed by infrared spectroscopy. However, the
limited resolution of the diffraction experiments did not allowed
a thorough study of the monoclinic distortion and the metallic phase
was not confirmed. 

In the present paper we report \textit{in situ }high pressure XAS
experiments for $YNiO_{3}$ up to 19~GPa. The expected IM transition
is finger-printed by the XAS near-edge features around 14~GPa. Around
that pressure we do not observe any modification in the $NiO_{6}$
geometry. The changes are mostly coming from a rearrangement of the
octahedra that leads to a straightening of the superexchange angle.
The subsequent conclusion is that the delocalization of the $e_{g}$
electrons is due to band effects, enabled in the orthorhombic phase
by favorable orbital alignments. Our results show
that XAS spectroscopy may provide the essential information about
octahedral links, hindered in scattering techniques by dynamical fluctuation
in the short range scale. We show that the occurrence of the MI transition
is not related to the exact local geometry of the $NiO_{6}$ octahedra
or charge disproportionation. It essentially depends on the middle
range organization. This confirms the paramount importance of the
octahedral tilting in the physics of the nickel perovskites, and in
a broader perspective, in correlated electron systems. 

\section{Experimental }

The pressure dependent XAS measurements at the Ni K-edge (8345~eV)
were performed at the dispersive XAS beamline \cite{Tolentino-PS05,Cezar-JSR10}
of the \textit{\emph{LNLS}} (Laboratório Nacional de Luz Síncrotron,
Campinas, Brazil). A fine grained high quality $YNiO_{3}$ polycrystalline
powder sample \cite{Alonso-PRB01} was loaded in the 125~$\mu m$
diameter hole of an iconel gasket mounted on 2.4 mm thick diamond
anvils, with a cullet of  300~$\mu m$. The spectra were measured
using a Si(111) bent crystal monochromator that was selecting a band-pass
of about 800~eV from the white beam and focusing it into a Gaussian
spot of $150\,\mu m$ -FWHM at the sample position. The gasket cut
the tails of the Gaussian beam. However, the beam position for each
energy was laterally dispersed in less than 30~$\mu m$, such that
the full energy band-pass were transmitted through the gasket with
almost the same photon flux. The reference flux, $I_{0}$, was measured
through a flat piece of glass in order to simulate the average attenuation
without introducing new features in the spectra. At each pressure,
the cell was realigned at the optical focus. In dispersive XAS there
is no mechanical movement of the optics during data collection and
the whole spectrum is observed at once. It then is possible to screen
the cell orientations in a relatively short time, to find an optimal
position. The remaining glitches were then removed by small rotations
of the cell with respect to the polychromatic X-ray beam. Using a
gas membrane-driven mechanism, the pressure was increased by steps
of about 2~GPa, measured using the ruby method\cite{Mao-JGR86}.
Pressures up to 20~GPa were applied using silicone oil as pressure
medium. Above 15~GPa the non-hydrostatic components of this medium
may lead to pressure deviations up to 10\% over a diameter of 150~$\mu m$.
The precision on the pressure is then around 0.5~GPa up to 10 GPa.
Above this value, an error bar of 10\%, associated to the non-hydrostaticity,
is estimated for the absolute pressure scale. Since XAS probes the
effective average contribution from all grains randomly oriented over
the pressure gradient in the sample area and due to the order of magnitude
of the effects (of the order of $10^{-2}$), XAS is not very sensitive
to a limited non-hydrostaticity. 

EXAFS (Extended X-ray Absorption Fine Structure) data free of Bragg
peaks were collected up to about 10.5$\textrm{\AA}^{-1}$. This data
range limits the minimum difference between two close distances that
can be resolved in the EXAFS analysis to the $\Delta R\simeq0.15\lyxmathsym{\AA}$.
In the $YNiO_{3}$ under ambient conditions the average Ni-O distance
is $1.994\,\lyxmathsym{\AA}$ and the separation between the largest
and shortest distance is only $0.039\,\lyxmathsym{\AA}$\cite{Alonso-PRB01}.
The restricted EXAFS k-range ($\simeq$10 $\mathring{A}^{-1}$) characterizes
then a low-resolution study in R-space. The Ni-O bond lengths are
not distinguishable and bond length differences within the two \emph{Ni}
sites appears as a static disorder contribution to the total bond
length dispersion\cite{Piamonteze-PRB05a}. The EXAFS analysis gives
only the average distance. The limited k-range broadens the Ni-O peak
but does not change its average position. Shifts in this average distance
with increasing pressure can be tracked precisely, provided that the
analysis procedure is strictly the same for the whole series of measurements.
The data were analyzed using the Athena/Artemis package\cite{Ravel-JSR05}.
The EXAFS oscillations were extracted following the standard procedure
of the Athena code. The signal corresponding to the oxygen coordination
shell was selected by Fourier filtering and the structural parameter
were deduced from fitting procedures. In the fitting procedure the
number of free parameters was limited by the useful range and the
interval corresponding to the selected signal in the real space. After
a first screening of the individual data sets, all data were fitted
together using the option of multiple set analysis. This option increases
the number of independent points over the number of free parameters.
By setting the same (fitted) origin to the k-scale, it reduces substantially
the uncertainty on the relative decrease of the Ni-O distances with
pressure. The number of neighbors and amplitude factor were fixed.
The only amplitude parameter left for each data set is the Debye-Waller
term $\sigma$, that describes the dispersion of the Ni-O bond lengths.
The changes in Ni-O bond length $\delta R_{Ni-O}$ and the bond length
dispersion$\sigma$ are obtained with error bars around $0.005\lyxmathsym{\AA}$
and $0.02\lyxmathsym{\AA}$, respectively. 

For the XANES (X-ray Absorption Near Edge Structure) range the normalization
procedure is the same for all pressures. In order to avoid spectra
deformation we limit the data handling to simple operations: normalization
consists essentially in the subtraction of a straight line for background
and to a scaling of the edge jump to 1 in the range 150-250~eV above
the edge. Variations in the straight line subtraction and the exact
position of the normalization zone affect to a small extent the overall
shape of the spectra but very little the relative intensity of the
structures. Due to the monochromator bandwidth, detector spatial resolution
and the core-hole lifetime, the experimental resolution for the XANES
experiments is around 1.5 eV. However, a measure of the shift of the
edge energy with pressure is not limited by the experimental resolution
but by the stability of the spectrometer and counting statistics.
In a dispersive XAS setup the monochromator does not move allowing
a very high energy stability. During the whole experiment, a sharp
Bragg peak outside the analysis range allowed us to verify that the
energy stability was better than 50 meV, in accordance with previous
measurements\cite{Cezar-JSR10}.

The XANES features were compared to \emph{ab initio} full multiple
scattering calculations using the FDMNES code \cite{Joly-PRB01} for
Ni-centered clusters and with atomic positions given by reported crystallographic
structures\cite{Garcia-Munoz-PRB92,Alonso-PRB01}. 

\begin{figure}
\includegraphics[scale=0.4]{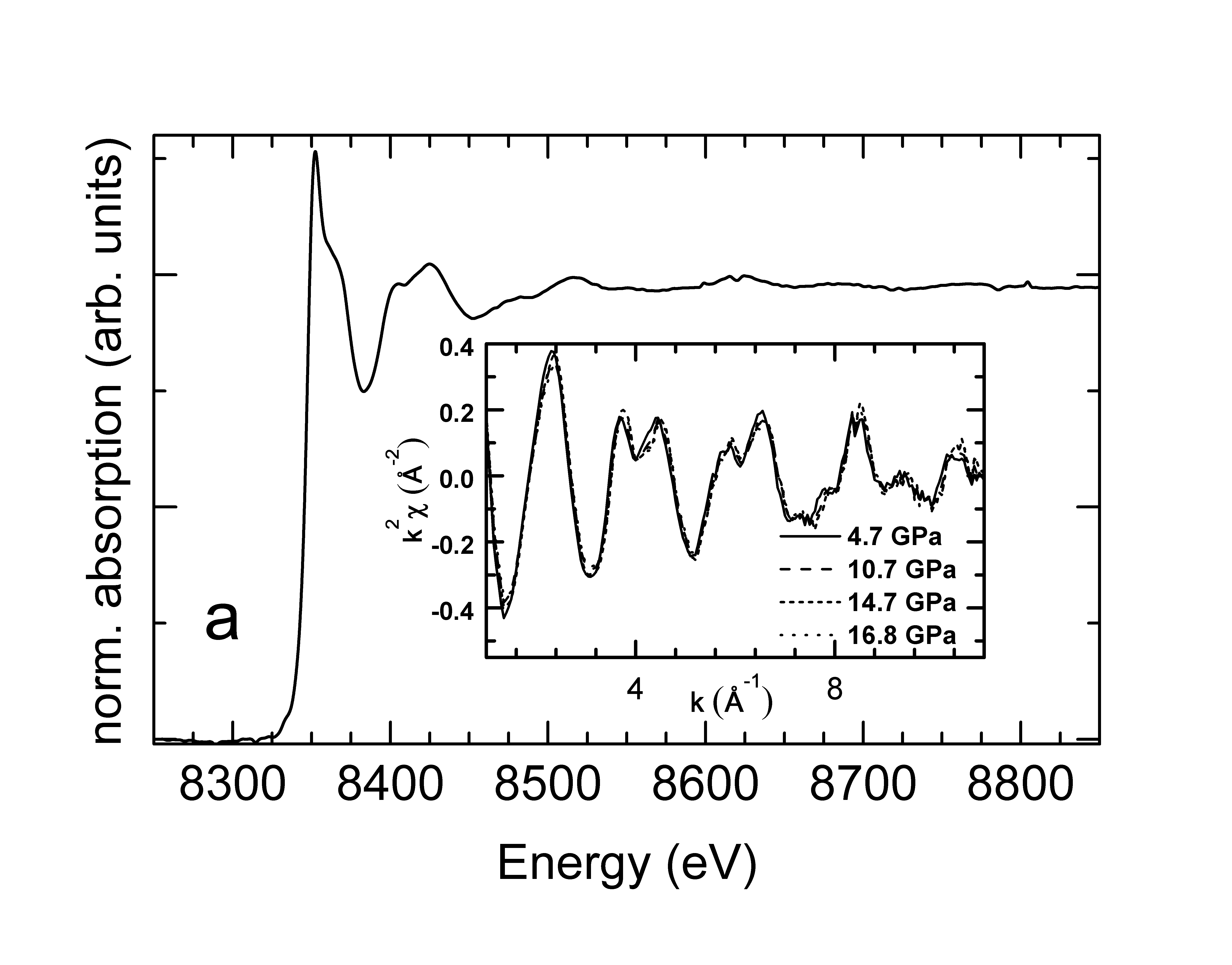}

\includegraphics[scale=0.4]{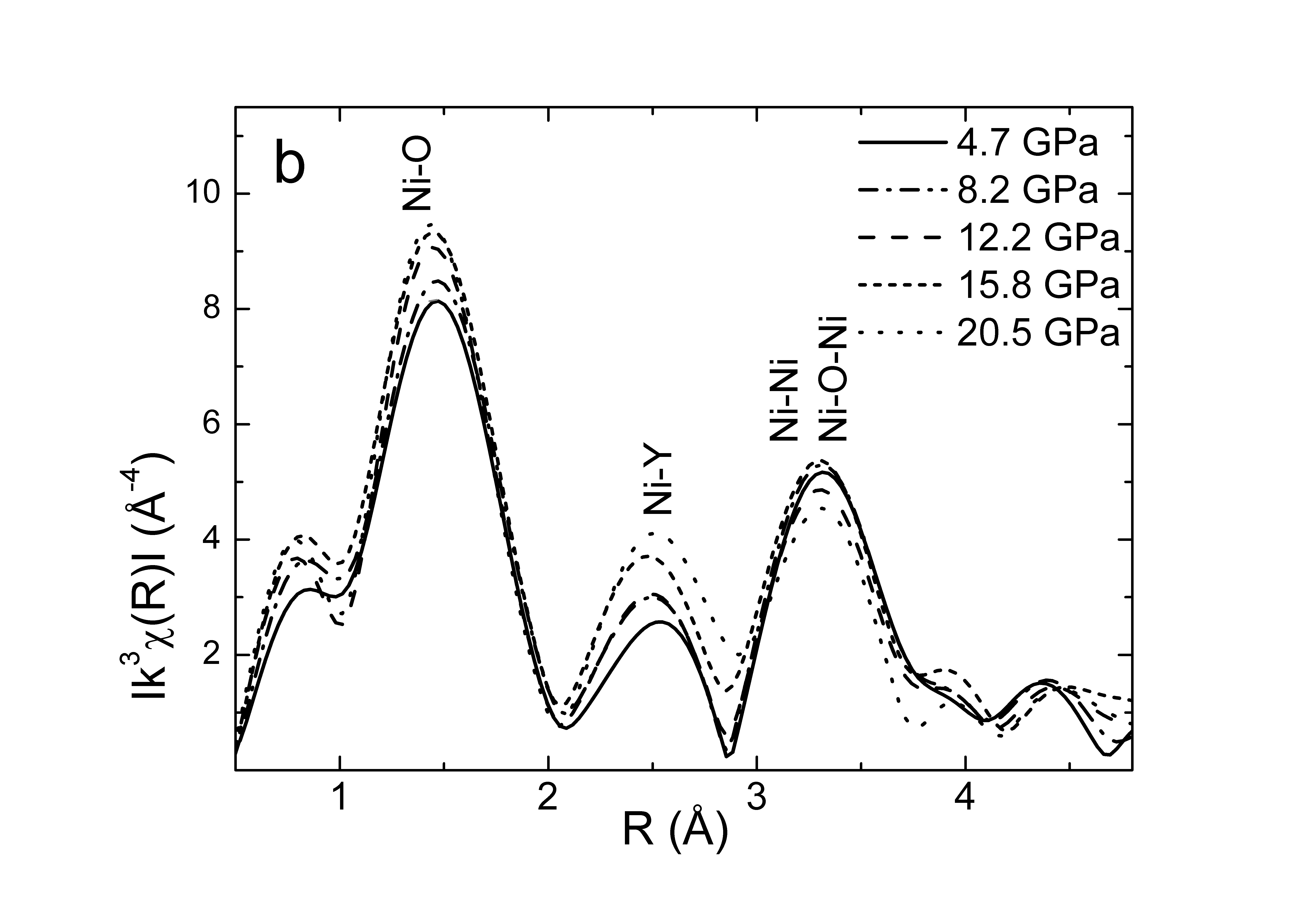}\caption{a: typical Ni K-edge XAS spectrum of $YNiO_{3}$. Inset: EXAFS signal
in the available k-range for different pressures. b Fourier transform
modulus of the EXAFS oscillations for selected pressures. }
\end{figure}

\section{results}

Figure 1-a shows the absorption spectrum at ambient pressure and the
EXAFS signal for selected pressures (inset). Figure 1-b gives the
modulus of the Fourier transform (FTM) of these signals. This representation
gives the pseudo-radial distribution function (not corrected for EXAFS
phase shifts) around the average Ni atom. The position of the most
prominent peak around $1.5\textrm{\,\AA}$ corresponds to the average
Ni-O bond length. The next two peaks correspond essentially to the
Ni-Y bondings and to the Ni-Ni backscattering and Ni-O-Ni multiple
scattering, respectively. These contributions reveal the sensitivity
of EXAFS to the next-nearest neighbors geometry and, so, to the bond
angles among octahedra. 

The progressive shift of the first peak of the FTM towards shorter
distances (Fig.1-b) yields the contraction $\Delta R$ in the average
Ni-O bond lengths. Figure 2 shows the pressure dependence of the Ni-O
distance contraction, obtained from the quantitative analysis after
selection of that peak. Between 2 and 13~GPa, $\Delta R$ decreases
almost linearly with the applied pressure. The contraction $\Delta R$
up to 13~GPa gives $\thickapprox0.033\,\textrm{\AA}$. Such a contraction
represents a $NiO_{6}$ relative volume decrease of $5.0\%$. In the
same range the unit cell volume measured by diffraction also drops
by $5.0\%$ \cite{Garcia-Munoz-PRB04}. The monotonic decrease of
the Ni-O average distance stops at around 13~GPa. In a transient
pressure range (13~GPa < P < 16~GPa), the average Ni-O bond length
and the distance dispersion remain almost unchanged. Above 16~GPa,
average Ni-O bond length resumes shrinking. The total contraction
up to 20~GPa is $\thickapprox0.043\,\textrm{\AA}$, or $6.6\%$ in
terms of volume decrease. The increase of the amplitude of the first
peak (Fig.1-b) corresponds to a decrease in the Ni-O bond length dispersion.
The largest increase occurs from 7 to 13~GPa. Within this range,
it the average $NiO_{6}$ units tend toward a partial symmetrization.
The evolution of the Debye-Waller term (Fig.2, inset) shows that indeed
the largest decrease in the bond dispersion occurs from 7 to 13~GPa,
confirming this qualitative outcome. 

The strongest modifications in the medium-range structures take place
within the pressure range 13-16~GPa. As seen in figure 1-b, the intensity
of the Ni-Y peak increases while those of the Ni-Ni and Ni-O-Ni peaks
decrease. These modifications point out inter-octahedra rearrangements.
Qualitatively, the evolution of position of the Ni-Y peak towards
low R is easily interpreted in terms of shortening of the Ni-Y distances.
The \textquotedblleft{}shift back\textquotedblleft{} towards larger
distances observed at pressures higher than 15~GPa does not fit with
such a simple scheme. One may note that already at 15~GPa the peak
is substantially enlarged on the high R side. As this enlargement
occurs at the pressure for which the monoclinic to orthorhombic transition
is expected, it could be a finger-print of the phase transition, and
consequently of some inter-octahedra rearrangement.

\begin{figure}
\includegraphics[scale=0.4]{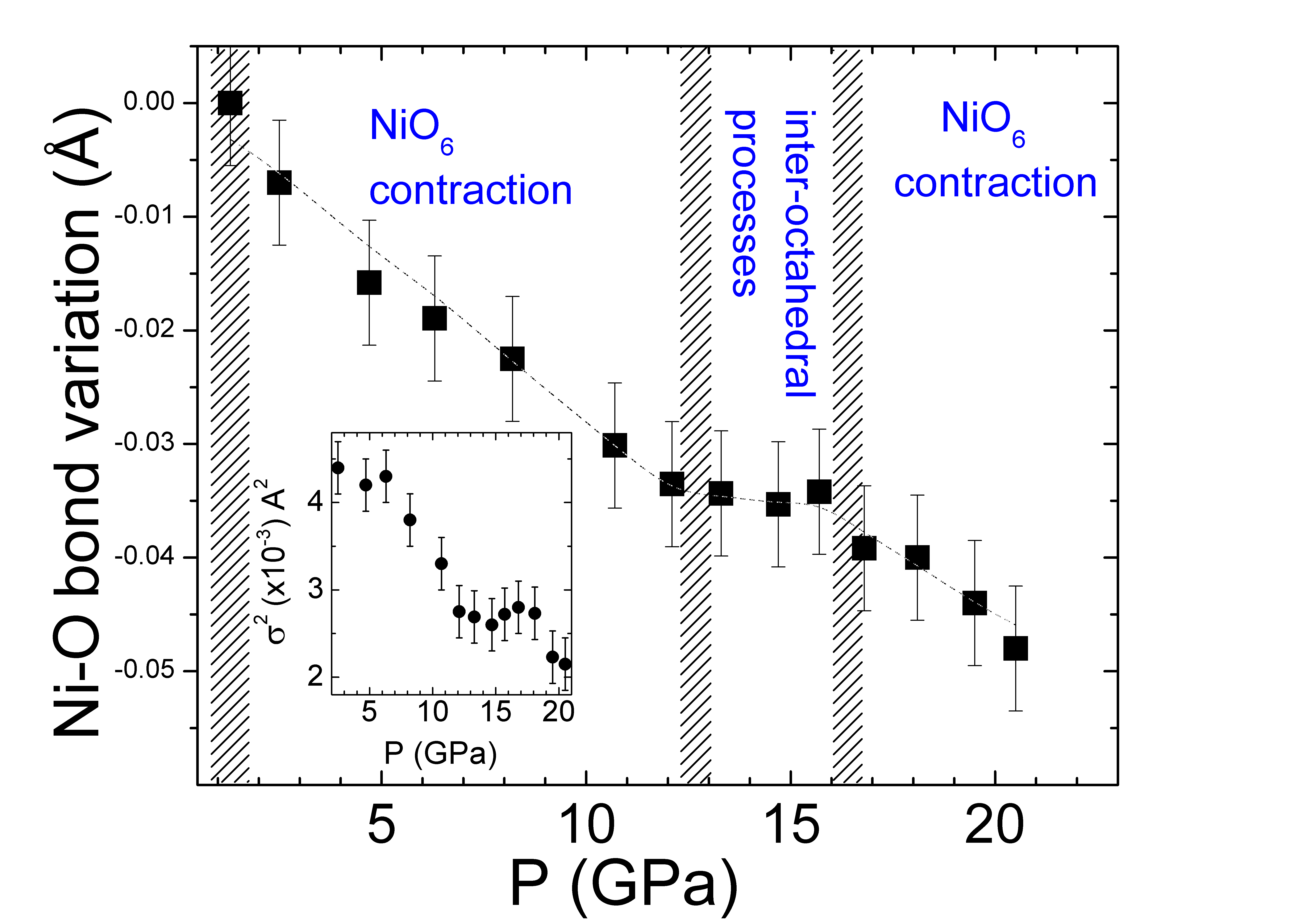}\caption{Pressure dependence of the average Ni-O bond length contraction $\Delta R$
obtained from the EXAFS analysis. The line is just a guide for the
eyes. Vertical lines separate different regions discussed in the text.
Inset : $\sigma^{2}$ as a function of the pressure.}
\end{figure}

Figure 3 shows the $NiK$-edge XANES spectra of $YNiO_{3}$ at selected
pressures. The pre-edge feature A, which probes the Ni 3d states hybridized
with Ni 4p states, does not change within the experimental precision.
Two main changes are observed when the applied pressure is increased:
a shift of the absorption threshold and subtle modifications of the
spectral features B and C. 

Edge shifts are primarily associated to changes in the formal valence.
However, as here the Ni keeps the form $Ni^{3+}$, these shifts have
essentially a structural origin. As coordination bond lengths decrease,
the band energy will increase as a result of the higher overlap of
the electrons density of neighboring atoms. This increased overlap
is less significant for localized 1s levels than for 4p delocalized
ones and, consequently, the absorption edge shifts towards higher
energies. For small shifts the bond length reduction ($\delta R$)
and the associated edge shift ($\delta E$) are almost linearly related\cite{SouzaNeto-PRB04,Ramos-PRB07}.
However, due to the mixture of close\emph{ Ni} bonds such linear relationship
does not strictly apply in $YNiO_{3}$, and the evolution of the edge
yields only a qualitative analysis. 

The edge shift $\delta E$ is measured by the position of the derivative
maximum. The observed shift rate, $\delta E/\delta P$, as a function
of applied pressure is not constant over the whole pressure range
(Fig.3- inset). Up to 13~GPa, $\delta E$ varies linearly with the
applied pressure and $\delta E/\delta P$ is about 60~meV/GPa. Within
the range 13~GPa < P < 16~GPa, the edge position position shifts
at a much slower rate. A new increase in $\delta E/\delta P$ above
16~GPa indicates the recovering of a regime of bond contraction.
All these outcomes are in full qualitative agreement with the EXAFS
analysis. 

\begin{figure}
\includegraphics[scale=0.4]{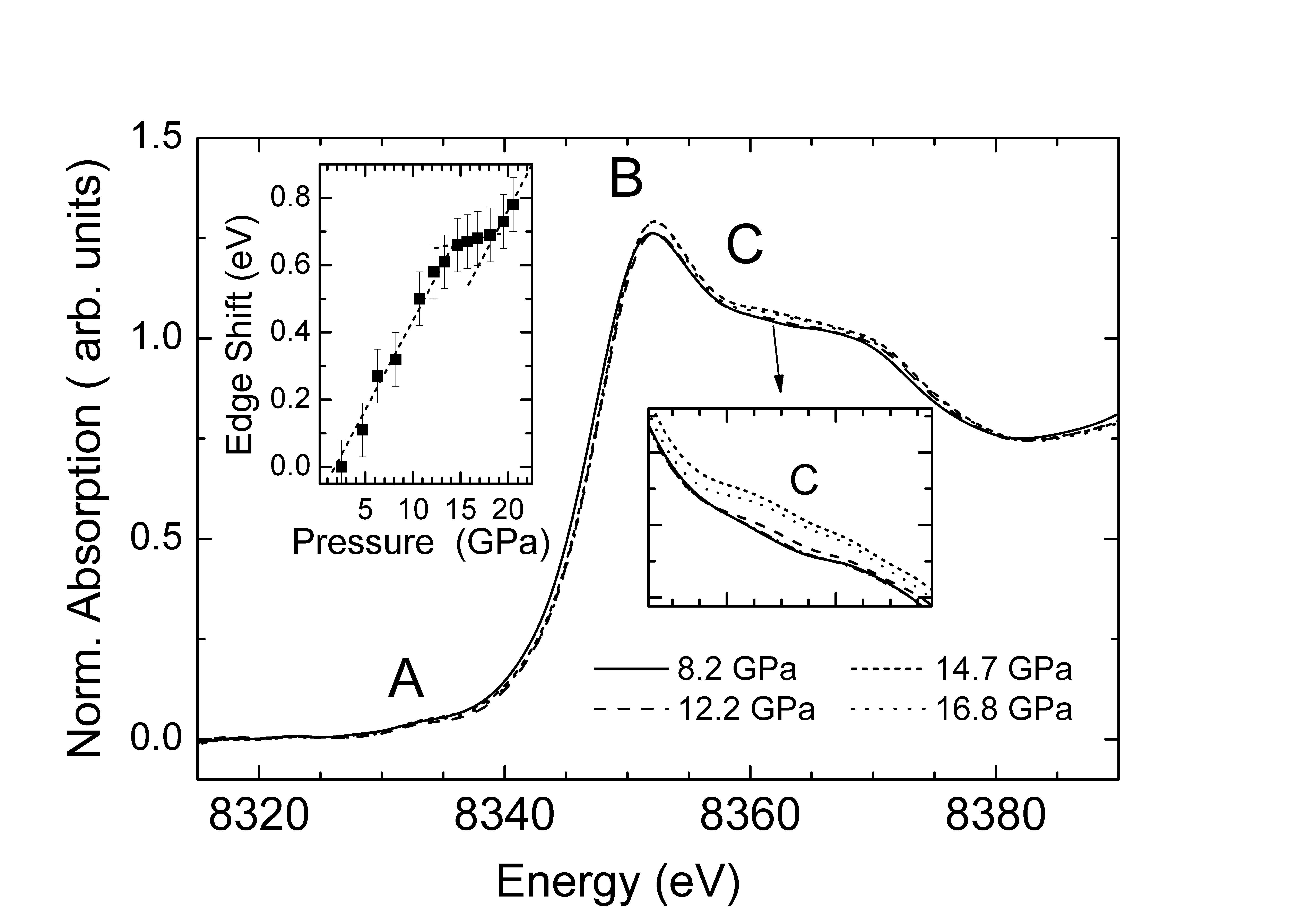}\caption{XANES spectra in $YNiO_{3}$ for selected representative pressures.
A is the pre-edge, B the main edge structure and C the shoulder associated
with middle range effects (see text). Lower inset : zoom on the C
feature. Upper inset: maximum of the derivative at the edge as a function
of the pressure. }
\end{figure}

Within the range 13~GPa < P < 16~GPa, a subtle raise of a shoulder
C about 10 eV above the main structure B is observed. The emergence
of similar shoulder was reported in $NdNiO_{3}$~ and $PrNiO_{3}$~\cite{Acosta-PRB08,Medarde-PRB92}.
In these studies it has been clearly associated to thermal induced
metallization. These authors relate the appearance of the shoulder
C to changes in the local atomic environment around Ni at the \textit{IM}
transition. The physical origin of this spectral feature will be discussed
below. Nevertheless, we can infer that in the present case the shoulder
C must also be associated to metallization. We note that its appearance
takes place in a narrow range around 14~GPa, where metallization
is indeed expected\cite{Garcia-Munoz-PRB04}, but also where the EXAFS
analysis shows that there is no significant change in the Ni-O bond
lengths. 

\begin{figure}
\includegraphics[scale=0.4]{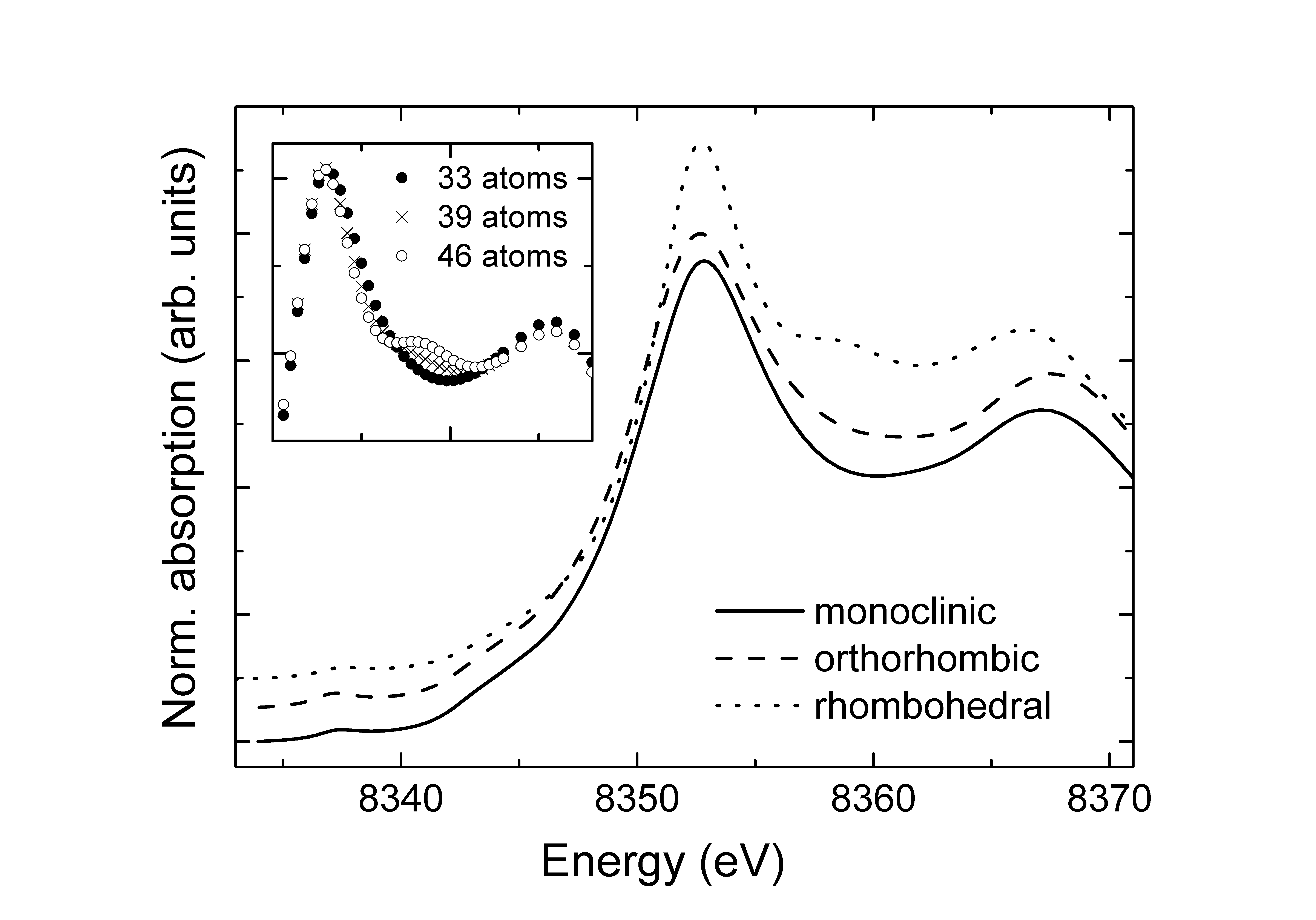}\caption{\textit{ab initio} simulations of the XANES spectra for a Ni centered
61 atoms-cluster of $YNiO_{3}$ in the monoclinic (plain line), orthorhombic
(dashed line) and $LaNiO_{3}$-like rhombohedral (dotted line) structures
Inset : simulations for the rhombohedral structure and different cluster
sizes . }
\end{figure}

We simulated the Ni K edge XANES spectra for $YNiO_{3}$ using\emph{
ab initio} full-multiple scattering calculations\cite{Joly-PRB01}
with atomic positions given by crystallographic structures reported
by neutron diffraction\cite{Alonso-PRB01,Garcia-Munoz-PRB92}. We
observed that the simulations using the monoclinic or the orthorhombic
crystallographic structures for $YNiO_{3}$ lead to identical spectra,
and do not show the expected C shoulder (Fig.4). Similar comparison
of the calculated spectra in $PrNiO_{3}$, for the reported orthorhombic
structure in the metallic and insulator phases, led Acosta and coworkers\cite{Acosta-PRB08}
to deduce that the changes in the local atomic structure are not reflected
in the average crystalline structure. Our results clearly support
this view. In addition, the absence of the structure C in our calculation
based on the monoclinic structure shows that this feature is not related
to the existence of two Ni sites. This suggests that C comes from
some characteristics of the local organization beyond the coordination
shell. 

To gather elements about the origin of this feature, we went further
simulating the XANES spectra based on a rhombohedral structure. Even
if not reported for $YNiO_{3}$, rhombohedral structure corresponds
to the most symmetric structure in the $RNiO_{3}$ series. Importantly,
in this structure the tilt angles among adjacent octahedra are quite
different compared to those found in monoclinic and orthorhombic $YNiO_{3}$.
This provides a track to interpret our data. The cluster was built
from cell parameters deduced from those of $LaNiO_{3}$ substituting
La by Y in the cell. In order to facilitate the comparison with the
experimental features in~$YNiO_{3}$, we use an isometric expansion
of these parameters keeping the cell volume that of the $YNiO_{3}$.
In this rombohedral structure the nickel atoms have a unique regular
octahedral site and the superexchange Ni-O-Ni angle is 165 degrees.
As for all previous calculations, the cluster contained 61 atoms.
The shoulder C is present around 10 eV above B (Fig.4). We then performed
simulations using the same $LaNiO_{3}$-rhombohedral structure but
with decreasing cluster size (Fig.4 inset). We checked that the C
shoulder does not correspond to a specific path of multiple scattering
within the octahedron. This structure emerges when the cluster considered
in the simulations includes at least 46 atoms (cluster size $\approx4.8\,\textrm{\AA}$),
i.e. half of the neighboring oxygen's octahedra.

\section{Discussion and conclusion }

We used X-ray absorption spectroscopy to describe the evolution of
the local and middle range order under applied pressure. A first regime
of monotonic contraction and symmetrization of the average $NiO_{6}$
units takes place up to 13 GPa. It stops in a limited pressure domain
around 14 GPa, before resuming above 16~GPa. In this narrow pressure
range, crystallographic modifications basically occur in the medium/long
range, not in the $NiO_{6}$ octahedron, whereas the evolution of
the near-edge XAS features can be associated to metallization. One
important outcome in our experiment is the emergence of the shoulder
C in the XANES around 14~GPa. Such emergence is reproduced in the
simulations when the cluster radius reaches a critical value of $\approx4.8\,\textrm{\AA}$.
It is then not related to the actual very local geometry or symmetrization
of the Ni sites but to middle range effect. This view is supported
by the experimental outcome that the appearance of the shoulder C
occurs in a narrow pressure range around 14~GPa where no significant
modifications in the average $NiO_{6}$ octahedron are observed, whereas
significant modifications seem to take place in the middle range order.
We should point out that in the experimental studies reported in $PrNiO_{3}$
and $NdNiO_{3}$\cite{Acosta-PRB08,Medarde-PRB92}, the shoulder C
rising at the IM transition is more intense than in our study. We
attributed that to our limited resolution, where a compromise had
to be found between good resolution for XANES and energy range for
EXAFS. As a consequence, the absolute intensity of the experimental
features C is not well reproduced in the simulations.

The main effect of pressure is to decrease bond lengths and straighten
bond angles leading to an increased bandwidth ($W$) and to a correlated
decrease in $T_{IM}$ \cite{Harrison-1980,Zhou-PRB00}. $T_{IM}$
goes down to room temperature at 14~GPa owing to the increased bandwidth,
which changes the ratio W/U, where U is the on-site Coulomb energy.
U is not expected to significantly modify around 14~GPa because it
concerns much more localized 3d orbitals and, in addition, the bond
length modification is very small. Although electron-electron correlation
(U) represents the main energy in these strong correlated systems,
the increase in the band width (W) seems to control the electronic
transition. 

The presence or not of the shoulder C around 15 eV above the absorption
threshold in the XANES spectra of rare earth nickelate turns out to
be related to the ability of the Ni 3d $e_{g}$ and O 2p orbitals
to overlap to form a band. Besides the Ni-O bonding interaction this
ability is strongly related to the Ni-O-Ni angle. The Ni-O-Ni angle
crossover for the $IM$ transition at room temperature and pressure
in $RNiO_{3}$ is around 156 degrees\cite{Zhou-PRB04b,Catalan-PhaseTrans08}.\textit{
}In both the monoclinic and orthorhombic crystallographic structures
given by neutron diffraction the Ni-O-Ni angle is around 146 degrees
in $YNiO_{3}$, while this angle is around 157 degrees in $PrNiO_{3}$
and $NdNiO_{3}$. The XANES results are in perfect agreement with
the EXAFS outcome that around 14~GPa the reduction of the distance
within the octahedra slows down, while a significant reorganization
takes place for the Y and Ni next neighbors. The drop in the lattice
parameter c identified by Garcia-Munoz and coworkers at 14~GPa\cite{Garcia-Munoz-PRB04}
corresponds more likely to inter-octahedra rearrangements than to
a modification in the local geometry of the $NiO_{6}$ units.

In conclusion, pressure dependent X-ray absorption spectroscopy in
$YNiO_{3}$ shows that, after an initial shrinking, there are no significant
modifications in the average $NiO_{6}$ octahedra around 14~GPa,
where a signature of electronic delocalization is observed. This experimental
outcome, supported by \textit{ab initio} XAS calculations, provides
evidences of the bandwidth driven nature of the pressure induced IM
transition in the same way that it has been recently reported in $LaMnO_{3}$
\cite{Ramos-EPL11}. The electronic delocalization is essentially
synchronized with the opening of the Ni-O-Ni angle, an not to a sudden
modifications in the local geometry of the $NiO_{6}$ octahedra.
\begin{acknowledgments}
This work is supported by LNLS, CNPq and CNPq-CNRS agreement. JAA
and MJML thank the Spanish MICINN for funding the project MAT2010-16404. 
\end{acknowledgments}

\end{document}